\documentclass{iaus}
\newcommand\kms{{\rm\,km\,s^{-1}}}
\newcommand\msun{\rm\,M_\odot}
\newcommand\rsun{\rm\,R_\odot}

\usepackage{graphicx}

\title[On the origin of the hypervelocity runaway star HD\,271791] %% give here short title %%
{On the origin of the hypervelocity runaway star HD\,271791}

\author[V.V.Gvaramadze]   %% give here short author list %%
{V.V.Gvaramadze}
\affiliation{Sternberg Astronomical Institute, Moscow State University,
       Universitetskij Pr. 13, Moscow 119992, Russia\\ email: {\tt vgvaram@mx.iki.rssi.ru} }
\pubyear{2009}
\volume{266}  %% insert here IAU Symposium No.
\pagerange{1--6}
\jname{Star Clusters -- Basic Galactic Building Blocks throughout Time and Space}
\editors{R. de Grijs \& J. Lepine, eds.}
\begin{document}

\maketitle

\begin{abstract}
We discuss the origin of the runaway early B-type star HD\,271791
and show that its extremely high velocity ($\simeq 530-920 \, \kms$)
cannot be explained within the framework of the binary-supernova
ejection scenario. Instead, we suggest that HD\,271791 attained its
peculiar velocity in the course of a strong dynamical encounter
between two hard massive binaries or via an exchange encounter
between a hard massive binary and a very massive star, formed
through runaway mergers of ordinary massive stars in the dense core
of a young massive star cluster. 
\end{abstract}

\firstsection % if your document starts with a section,

\section{Introduction}

HD\,271791 is a massive ($11\pm 1 \, M_{\odot}$; Przibilla et al.
\cite{pr08}) extremely high-velocity runaway star with a Galactic
rest-frame velocity of $\simeq 530-920 \, \kms$ (Heber et al.
\cite{he08}), typical of the so-called hypervelocity stars (HVVs) --
the ordinary stars moving with peculiar velocities exceeding the
escape velocity of our Galaxy (Brown et al. \cite{br05}; Edelmann et
al. \cite{ed05}; Hirsch et al. \cite{hi05}). The existence of the
HVSs was predicted by Hills (\cite{hi88}), who showed that close
encounter between a tight binary system and the supermassive black
hole (BH) in the Galactic Centre could be responsible for ejection
of one of the binary components with a velocity of up to several
$1000 \, \kms$. Yu \& Tremaine (\cite{yu03}) proposed and additional
possible mechanism for production of HVSs based on the interaction
between a single star and a putative binary BH in the Galactic
Centre. It is therefore plausible that some HVSs were produced in
that way (Gualandris, Portegies Zwart \& Sipior \cite{gu05};
Baumgardt, Gualandris \& Portegies Zwart \cite{ba06}; Levin
\cite{le06}; Sesana, Haardt \& Madau \cite{se06}; Ginsburg \& Loeb
\cite{gi06}; Lu, Yu \& Lin \cite{lu07}; L\"{o}ckmann \& Baumgardt
\cite{lo08}). At present, however, proper motion measurements are
not available for all but one of the known HVSs so that it is
impossible to unambiguously associate their birthplace with the
Galactic Centre. HD\,271791 is the only HVS with measured proper
motion and all measurements show that this star was ejected from the
periphery of the Galactic disc (at a galactocentric distance of
$\geq 15$ kpc; Heber et al. \cite{he08}).

There are two possible alternative explanations of the origin of
HVSs. The first one is that the HVSs attain their high peculiar
velocities in the course of strong dynamical three- or four-body
encounters in young and dense star clusters located in the Galactic
disc (Gvaramadze \cite{gv06a}, \cite{gv07}; Gvaramadze, Gualandris
\& Portegies Zwart \cite{gva08}, \cite{gva09}) or in the Large
Magellanic Cloud (Gualandris \& Portegies Zwart \cite{gu07}). The
second one was proposed by Abadi, Navarro \& Steinmetz
(\cite{ab09}). According to these authors, some HVSs could originate
from tidal disruption of dwarf galaxies during their close passage
near the Milky Way.  The young age of HD\,271791 of $25\pm 5$ Myr
(Przybilla et al. \cite{pr08}) is inconsistent with the second
possibility since there are no indications of a recent encounter
between a dwarf satellite with the Milky Way. So, we are left with
the first one. Before discussing it in Section\,4, we consider a
proposal by Przybilla et al. (\cite{pr08}) that HD\,271791 attained
its peculiar velocity in the course of disintegration of a close
massive binary system following the supernova (SN) explosion.

\section{HD\,271791: former secondary in a massive binary system}

The spectral analysis of HD\,271791 by Przybilla et al.
(\cite{pr08}) revealed that the Fe abundance in its atmosphere is
subsolar and that the $\alpha$-process elements are enhanced. The
first finding is consistent with the origin of HD\,271791 in the
metal-poor outskirts of the Galactic disc, while the second one
suggests that this star was a secondary component of a massive tight
binary and that its surface was polluted by the nucleosynthetic
products after the primary star exploded in a SN. Przybilla et al.
(\cite{pr08}) believe that the binary-SN explosion could be
responsible not only for the $\alpha$-enhancement in HD\,271791 but
also for the extremely high space velocity of HD\,271791. Below, we
outline their scenario.

The large separation of HD\,271791 from the Galactic plane ($\simeq
10$ kpc) along with the proper motion measurements (Heber et al.
\cite{he08}) implies that the time-of-flight of this B2III star is
comparable to its evolutionary age, which in turn implies that the
star was ejected within several Myr after its birth in the Galactic
disc. The ejection event was connected with disruption of a massive
tight binary following the SN explosion. The original binary was
composed of a primary star of mass of $\geq 55 \, \msun$ and an
early B-type secondary (HD\,271791), so that the SN explosion and
the binary disruption occurred early in the lifetime of HD\,271791.
The system was close enough to go through the common-envelope phase
before the primary exploded in a SN. During the common-envelope
phase, the primary star lost most of its hydrogen envelope and the
binary became a tight system composing of a Wolf-Rayet star and an
early B-type main-sequence star. At the moment of SN explosion, the
mass of the primary star was $\leq 20 \, \msun$ and the binary
semimajor axis was $\sim 14 \rsun$ (that corresponds to the orbital
velocity of the secondary of $\leq 420 \, \kms$). The exploded star
expelled $\simeq 10 \, \msun$ of its mass while the remaining mass
collapsed to a $\leq 10 \, \msun$ BH. The SN explosion was
asymmetric enough to disrupt the system. Przybilla et al.
(\cite{pr08}) assumed that HD\,271791 was released at its orbital
velocity and that at the time of binary disruption the vector of the
orbital velocity was directed by chance along the Galactic rotation
direction. The first assumption is based on the wide-spread
erroneous belief that runaways produced from a SN in a binary system
have peculiar velocities comparable to their pre-SN orbital
velocities. The second assumption is required to explain the
difference between the assumed space velocity from the binary
disruption and the Galactic rest-frame velocity of HD\,271791
(provided that the latter is on the low end of the observed range
$530-920 \, \kms$).

In the next section, we discuss the conditions under which the
secondary star could be launched into free flight at a velocity
equal to its pre-SN orbital one.

\section{Binary-supernova ejection scenario}

\begin{figure}[b]
\includegraphics[width=0.9\columnwidth,angle=0]{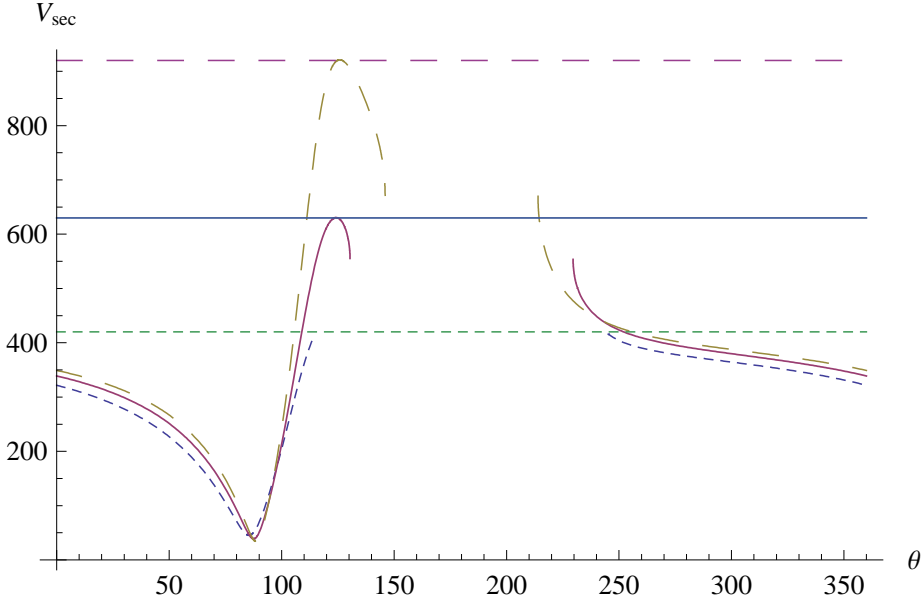}
\begin{center}
\caption{The peculiar velocity
of the secondary star (HD\,271791) as a function of the angle, $\theta$,
between the kick vector and the direction of motion of the exploding
star and the magnitude of the kick [$750 \, \kms$ (short-dashed line),
$1000 \, \kms$ (solid line), $1200 \, \kms$ (long-dashed line)]. The discontinuities in the
curves correspond to a range of angles $\theta$ for which the system remains
bound. The horizontal solid line indicates the Galactic rest-frame
velocity of HD\,271791 of $630 \, \kms$, corresponding to the "best"
proper motion given in Heber et al. (\cite{he08}). The horizontal short- and long-dashed
lines indicate, respectively, the orbital velocity of HD\,271791 of $420 \, \kms$
[suggested by the scenario of Przybilla et al. (\cite{pr08})] and the maximum
possible Galactic rest-frame velocity of HD\,271791 of $920 \, \kms$ (Heber et al.
\cite{he08}). See text for details.}
\end{center}
\end{figure}

One of two basic mechanisms producing runaway stars is based on a SN
explosion in a massive tight binary system (Blaauw \cite{bl61}).
After the primary star exploded in a SN, the binary system could be
disintegrated if the system lost more than half of its pre-SN mass
(Boersma \cite{bo61}) and/or the SN explosion was asymmetric (so
that the stellar remnant, either a neutron star (NS) or a BH,
received at birth a kick velocity exceeding the escape velocity from
the system; Stone \cite{st82}; Tauris \& Takens \cite{ta98}).

In the case of binary disruption following the symmetric SN
explosion, the stellar remnant is released at its orbital velocity,
while the space velocity of the secondary star, $V_{\rm sec}$, is
given by (Boersma \cite{bo61}; Radhakrishnan \& Shukre \cite{ra85};
Tauris \& Takens \cite{ta98})
\begin{equation}
V_{\rm sec} = \sqrt{1-2{m_1 +m_2 \over m_1 ^2}} \, V_{\rm orb} \, ,
\end{equation}
where $m_1 > 2 + m_2 , m_1 =M_1 /M_{\rm co} , m_2 =M_2 /M_{\rm co}$,
$M_1$ and $M_2$ are the pre-SN masses of the primary and the
secondary stars, $M_{\rm co}$ is the mass of the compact object
formed in the SN explosion, $V_{\rm orb} = [GM_1 ^2 /(M_1
+M_2)a]^{1/2}$ is the orbital velocity of the secondary star, $G$ is
the gravitational constant and $a$ is the binary semimajor axis. It
follows from equation\,(2.1) that $V_{\rm sec} \simeq V_{\rm orb}
\simeq (GM_1 /a)^{1/2}$ if $M_1 >> M_2 , M_{\rm co}$.

The above consideration shows that the secondary star could achieve
a high peculiar speed if one adopts a large pre-SN mass of the
primary (i.e. $M_1 >> M_{\rm co}$). But, the stellar evolutionary
models suggest that the pre-SN masses of stars with initial masses,
from 12 to $120 \, \msun$ do not exceed $\sim 10-17 \, \msun$
(Vanbeveren, De Loore \& Van Rensbergen \cite{va98}; Meynet \&
Maeder \cite{me03}). Using these figures and assuming that the
pre-SN binary is as tight as possible (i.e. the secondary
main-sequence star is close to filling its Roche lobe), one can
estimate the maximum possible velocity achieved by a runaway star in
the process of binary disruption following the symmetric SN
explosion. Assuming that the SN explosion left behind a NS (i.e.
$M_{\rm co} =1.4 \, \msun$) and adopting $M_1 =10-17 \, \msun$, one
has that a $3 \, \msun$ secondary star could attain a peculiar
velocity of $\simeq 300-500 \kms$, while a $10 \, \msun$ star could
be ejected with a speed of $\simeq \, 350 \kms$.

Note that the pre-SN mass is maximum for stars with initial mass
$\simeq 20-25 \, \msun$ and $\geq 80 \, \msun$ [see Fig.\,6 of
Meynet \& Maeder (\cite{me03})]. In the first case, the SN explosion
leave behind a NS, while in the second one the stellar SN remnant is
a BH of mass $M_{\rm co} \geq 5 \, \msun$ (e.g. Woosley et al.
\cite{wo02}). The large separation of HD\,271791 from the Galactic
plane implies that this massive star was ejected very soon after its
birth in the Galactic disc. From this, it follows that to explain
the space velocity of HD\,271791 within the framework of the
binary-SN scenario one should assume that the primary was a
short-lived very massive star (Przybilla et al. \cite{pr08}). In
this case, the stellar SN remnant is a BH and the SN ejecta is not
massive enough to cause the disruption of the binary system.

According to Przybilla et al. (\cite{pr08}), the pre-SN mass of the
primary star was $\leq 20 \, \msun$ (i.e. somewhat larger than the
maximum mass predicted by the stellar evolutionary models; see
above) and the SN explosion left behind a BH of mass $\leq 10 \,
\msun$ (comparable to the mass of the secondary star, HD\,271791),
i.e. the system lost less than a half of its mass. Thus, to disrupt
the binary, the SN explosion should be asymmetric. In this case, the
space velocities of the BH and HD\,271791 depend on the magnitude
and the direction of the kick imparted to the BH at birth (Tauris \&
Takens \cite{ta98}). To estimate $V_{\rm sec}$, one can use
equations\,(44)-(47) and (54)-(56) given in Tauris \& Takens
(\cite{ta98}). It follows from these equations that $V_{\rm sec}$ is
maximum if the vector of the kick velocity does not strongly deviate
from the orbital plane of the binary and is directed nearly towards
the secondary, i.e. the angle, $\theta$, between the kick vector and
the direction of motion of the exploding star is $\sim \theta
_{\ast} = \arccos (-v/w)$, where $v=[G(M_1 +M_2 )/a]^{1/2}$ is the
relative orbital velocity and $w$ is the kick velocity (see
Gvaramadze \cite{gv06b}).

Fig.\,1 shows how the direction and the magnitude of the kick affect
$V_{\rm sec}$. The three solid lines represent $V_{\rm sec}$
calculated for the binary parameters suggested by Przybilla et al.
(\cite{pr08}) and three kick magnitudes of $750 \, \kms$
(short-dashed), $1000 \, \kms$ (solid line) and $1200 \, \kms$
(long-dashed line). One can see that to launch HD\,271791 at its
pre-SN orbital velocity $V_{\rm orb} \simeq 420 \, \kms$ the kick
imparted to the BH should be at least as large as $750 \, \kms$. In
fact, the kick magnitude should be much larger since for kicks
$\simeq 750 \, \kms$ the kick direction must be carefully tuned (see
Fig\,1), i.e. $\theta$ should be either $\simeq 115^{\circ}$ or
$\simeq 245^{\circ}$ (note that for $115^{\circ} \leq \theta \leq
245^{\circ}$ the binary system remains bound). The even larger kicks
of $\geq 1000$ and $\geq 1200 \, \kms$ are required to explain the
Galactic rest-frame velocities of HD\,271791 of $630$ and $920 \,
\kms$ [corresponding, respectively, to the "best" and the maximum
proper motions given in Heber et al. (\cite{he08}); see also
Przybilla et al. (\cite{pr08})]. Although one cannot exclude a
possibility that BHs attain a kick at birth, we note that there is
no evidence that the kick magnitude could be as large as required by
the above considerations (e.g. Nelemans et al. \cite{ne99}).

Thus, we found that to explain the peculiar velocity of HD\,271791
the magnitude of the kick attained by the $\leq 10 \, \msun$ BH
should be unrealistically large ($\geq 750 \, \kms$), that makes the
binary-SN ejection scenario highly unlikely (cf. Gvaramadze
\cite{gv07}; Gvaramadze \& Bomans \cite{gv08}). Nevertheless, some
authors still believe that ``HD 271791 is a plausible runaway star
produced by a supernova explosion in a massive binary system"
(Bromley et al. \cite{br09}).

\section{Dynamical ejection scenario}

The second basic mechanism responsible for the origin of runaway
stars is based on dynamical three- or four-body interactions in
dense stellar systems (Poveda et al. \cite{po67};
 Aarseth \cite{aa74};
Gies \& Bolton \cite{gi86}). Below, we discuss two possible channels
for producing high-velocity runaways within the framework of the
dynamical ejection scenario (cf. Gvaramadze \cite{gv09}).

The first possibility is that the high-velocity runaways originate
through the interaction between two massive hard binaries (Mikkola
\cite{mi83}; Leonard \& Duncan \cite{le90}). The runaways produced
in binary-binary encounters are frequently ejected at velocities
comparable to the orbital velocities of the binary components
(Leonard \& Duncan \cite{le90}) and occasionally they can attain a
velocity as high as the escape velocity, $V_{\rm esc} =(2GM_{\ast}
/R_{\ast} )^{-0.5}$, from the surface of the most massive star in
the binaries (Leonard \cite{le91}). For the upper main-sequence
stars with the mass-radius relationship (Habets \& Heintze
\cite{ha81}), $R_{\ast} =0.8(M_{\ast} /\msun )^{0.7} \rsun$, where
$R_{\ast}$ and $M_{\ast}$ are the stellar radius and the mass, one
has $V_{\rm esc} \simeq 700 \, \kms (M_{\ast}/\msun)^{0.15}$, so
that the ejection velocity could in principle be as large as $\simeq
1100-1200 \, \kms$ if the binaries contain at least one star of mass
of $20-40 \, \msun$.

To reconcile this ejection scenario with the presence of
nucleosynthetic products in the atmosphere of HD\,271791, one should
assume that (i) HD\,271791 was a secondary component of one of the
binaries involved in the encounter, and (ii) by the moment of the
encounter, the binary containing HD\,271791 has experienced
supernova explosion and remained bound [i.e. the stellar supernova
remnant (BH) received a small or no kick at birth]. The requirement
that HD\,271791 was a member of a post-supernova binary should also
be fulfilled in the second dynamical process discussed below.

The second possibility is that the high-velocity runaway stars
attain their peculiar velocities in the course of close encounters
between massive hard binaries and a very massive star (Gvaramadze
\cite{gv07}; Gvaramadze et al. \cite{gva09}), formed through runaway
collisions of ordinary massive stars in dense star clusters (e.g.
Portegies Zwart et al. \cite{po99}). A close encounter with the very
massive star results in a tidal breakup of the binary, after which
one of the binary components becomes bound to the very massive star
while the second one recoils with a high velocity, given by (Hills
\cite{hi88}):
\begin{equation}
V_{\infty} \simeq 500 \, \kms \, (M_{\rm VMS} /100 \, \msun )^{1/6}
({\it a}' /30\, \rsun )^{-1/2} ({\it M}_1 /10 \,\msun )^{1/3}\, ,
\end{equation}
where $M_{\rm VMS}$ is the mass of the very massive star and $a'$ is
the post-SN binary semimajor axis. It follows from equation\,(4.1)
that, to explain the peculiar velocity of HD\,271791 of $\simeq
400-600 \, \kms$, the mass of the very massive star should be $\geq
100-300 \, \msun$ [the first figure corresponds to the mass of the
most massive star formed in a `normal' way in a cluster with a mass
$M_{\rm cl} \simeq 10^4 \, \msun$ (Weidner, Kroupa \& Bonnell
\cite{we09})]\footnote{Simple estimates show that our Galaxy can
currently host about 100 star clusters with a mass $M_{\rm cl} \geq
10^4 \, \msun$ (Gvaramadze et al. \cite{gva08}).}. The above
estimates can be supported by the results of three-body scattering
experiments which showed that $\geq 3$ per cent of encounters
between hard massive binaries and a very massive star of mass of
$200-300 \, \msun$ produce runaways with $V_{\infty} \geq 500-600 \,
\kms$ (Gvaramadze et al. \cite{gva09}).

Note that the requirement that HD\,271791 was a member of a post-SN
binary does not contradict to our proposal that this star can attain
its high speed via a three-body encounter with a very massive star
(i.e. with the star more massive than the primary star in the
original binary). The merging of ordinary stars results in effective
rejuvenation of the collision product (e.g. Meurs \& van den Heuvel
\cite{me89}) so that the very massive star could still be on the
main sequence when the most massive ordinary stars start to explode
as SNe (e.g. Portegiez Zwart et al. \cite{po99}).

\section{Acknowledgements}
I am grateful to L.R.Yungelson for useful discussions, to the
Russian Foundation for Basic Research and the International
Astronomical Union for travel grants and to the Deutsche
Forschungsgemeinschaft for partial financial support.


\begin{thebibliography}{}

%
\bibitem[1974]{aa74} {Aarseth S.J.} 1974, {\it A\&A}, 35, 237
\bibitem[2009]{ab09} {Abadi, M.G., Navarro, J.F., Steinmetz, M.} 2009, {\it ApJ}, 691, L63
\bibitem[2006]{ba06} {Baumgardt, H., Gualandris, A., Portegies Zwart, S.} 2006, {\it MNRAS}, 372, 174
\bibitem[1961]{bl61} {Blaauw, A.} 1961, {\it Bull. Astron. Inst. Netherlands}, 15, 265
\bibitem[1961]{bo61} Boersma J., 1961, {\it Bull. Astron. Inst. Netherlands}, 15, 291
\bibitem[2009]{br09} {Bromley, B.C., Kenyon, S.J., Brown, W.R., Geller,
M.J.} 2009, {\it preprint} (arXiv0907.5567)
\bibitem[2005]{br05} {Brown W.R., Geller M.J., Kenyon S.J., Kurtz M.J.} 2005, {\it ApJ}, 622, L33
\bibitem[2005]{ed05} {Edelmann, H., Napiwotzki, R., Heber, U., Christlieb, N., Reimers, D.} 2005, {\it ApJ}, 634, L181
\bibitem[1986]{gi86} {Gies, D.R., Bolton, C.T.} 1986, {\it ApJS}, 61, 419
\bibitem[2006]{gi06} {Ginsburg, I., Loeb, A.} 2006, {\it MNRAS}, 368, 221
\bibitem[2007]{gu07} {Gualandris, A., Portegies Zwart, S.} 2007, {\it MNRAS}, 376, L29
\bibitem[2005]{gu05} {Gualandris, A., Portegies Zwart, S., Sipior, M.S.} 2005, {\it MNRAS}, 363, 223
\bibitem[2006a]{gv06a} {Gvaramadze, V.V.} 2006a, in {\it On the Present and Future of Pulsar Astronomy}, 26th meeting
of the IAU, Joint Discussion 6, 16-17 August, 2006, Prague, Czech
Republic, JD06, \# 25
\bibitem[2006b]{gv06b} {Gvaramadze, V.V.} 2006b, {\it A\&A}, 454, 239
\bibitem[2007]{gv07} {Gvaramadze, V.V.} 2007, {\it A\&A}, 470, L9
\bibitem[2009]{gv09} {Gvaramadze, V.V.} 2009, {\it MNRAS}, 395, L85
\bibitem[2008]{gv08} {Gvaramadze, V.V., Bomans, D.J.} 2008, {\it A\&A}, 485, L29
\bibitem[2008]{gva08} {Gvaramadze, V.V., Gualandris, A., Portegies Zwart, S.} 2008, {\it MNRAS}, 385, 929
\bibitem[2009]{gva09} {Gvaramadze, V.V., Gualandris, A., Portegies Zwart, S.} 2009, {\it MNRAS}, 396, 570
\bibitem[1981]{ha81} {Habets, G.M.H.J., Heintze, J.R.W.} 1981, {\it A\&AS}, 46, 193
\bibitem[2008]{he08} {Heber, U., Edelmann, H., Napiwotzki, R., Altmann, M., Scholz, R.-D.} 2008, {\it A\&A}, 483, L21
\bibitem[1988]{hi88} {Hills, J.G.} 1988, {\it Nat}, 331, 687
\bibitem[2005]{hi05} {Hirsch, H.A., Heber, U., O'Toole, S. J., Bresolin, F.} 2005, {\it A\&A}, 444, L61
\bibitem[1991]{le91} {Leonard, P.J.T.} 1991, {\it AJ}, 101, 562
\bibitem[1990]{le90} {Leonard, P.J.T., Duncan, M.J.} 1990, {\it AJ}, 99, 608
\bibitem[2006]{le06} {Levin, Y.} 2006, {\it ApJ}, 653, 1203
\bibitem[2008]{lo08} {L\"{o}ckmann, U., Baumgardt, H.} 2008, {\it MNRAS}, 384, 323
\bibitem[2007]{lu07} {Lu, Y., Yu, Q., Lin, D.N.C.} 2007, {\it ApJ}, 666, L89
\bibitem[1989]{me89} {Meurs, E.J.A., van den Heuvel, E.P.J.} 1989, {\it A\&A}, 226, 88
\bibitem[2003]{me03} {Meynet, G., Maeder, A.} 2003, {\it A\&A}, 404, 975
\bibitem[1983]{mi83} {Mikkola, S.} 1983, {\it MNRAS}, 203, 1107
\bibitem[1999]{ne99} {Nelemans, G., Tauris, T.M., van den Heuvel, E.P.J.} 1999, {\it A\&A}, 352, L87
\bibitem[1999]{po99} {Portegies Zwart, S.F., Makino, J., McMillan, S.L.W., Hut, P.} 1999, {\it A\&A}, 348, 117
\bibitem[1967]{po67} {Poveda, A., Ruiz, J., Allen, C.} 1967, {\it Bol. Obs. Tonantzintla Tacubaya}, 4, 86
\bibitem[2008]{pr08} {Przybilla, N., Nieva, M.F., Heber, U., Butler, K.} 2008, {\it ApJ}, 684, L103
\bibitem[1985]{ra85} {Radhakrishnan, V., Shukre, C.S.} 1985, in {\it Supernovae, Their Progenitors and
Remnants}, eds. G. Srinivasan \& V. Radhakrishnan, (Bangalore:
Indian Academy of Sciences), 155
\bibitem[2006]{se06} {Sesana, A., Haardt, F., Madau, P.} 2006, {\it ApJ}, 651, 392
\bibitem[1982]{st82} {Stone R.C.} 1982, {\it AJ}, 87, 90
\bibitem[1998]{ta98} {Tauris, T.M., Takens, R.J.} 1998, {\it A\&A}, 330, 1047
\bibitem[1998]{va98} {Vanbeveren, D., De Loore, C., Van Rensbergen, W.} 1998, {\it A\&AR}, 9, 63
\bibitem[2009]{we09} {Weidner, C., Kroupa, P., Bonnell, I.} 2009, {\it MNRAS}, in press (arXiv:0909.1555)
\bibitem[2002]{wo02} {Woosley, S.E., Heger, A., Weaver, T.A.} 2002, {\it RvMP}, 74, 1015
\bibitem[2003]{yu03} {Yu, Q., Tremaine, S.} 2003, {\it ApJ}, 599, 1129

\end{thebibliography}
\end{document}